# ESTIMATION OF THE ELECTRICAL AND THERMAL CONTACT RSISTANCES AND THERMOEMF OF "THERMOELECTRIC MATERIAL-METAL" TRANSIENT CONTACT LAYER DUE TO SEMICONDUCTOR SURFACE ROUGHNESS

*P.V. Gorskiy*

*Institute of Thermoelectricity of the NAS and MES of Ukraine, 1, Nauky str., Chernivtsi, 58029, Ukraine*

*The impact of semiconductor surface roughness on the electrical and thermal contact resistances and thermoEMF of "thermoelectric material (TEM)-metal" transient contact layer is studied theoretically. The distribution of "hollows" and "humps" on the rough surface is simulated by the "truncated Gaussian distribution". The impact of distribution parameters on the electrical contact resistance and thermoEMF of "thermoelectric material-metal" contact is studied. Specific numerical calculations and plotting is made for the case of bismuth telluride-nickel contact. It turned out that the electrical and thermal contact resistances and thermoEMF at low root-mean-square deviations of profile height nonmonotonically depend on the average profile height, however, as the distribution of "hollows" and "humps" approaches the uniform, they tend to certain asymptotic values. In so doing, both the thermal and electrical contact resistances and thermoEMF increase at high relative values of the average profile height and decrease and its low values.*

**Key words:** electrical contact resistance, thermoEMF, surface roughness, transient layer, elementary bars, average value, root-mean-square deviation.

## 1. Introduction

The previous authors published a number of theoretical works dedicated to calculations of "TEM-metal" electrical and thermal contact resistances [1-4]. These works dealt with the barrier and emission mechanisms of formation of "TEM-metal" electrical contact resistance, diffusion phonon scattering on surface irregularities as a mechanism of formation of "TEM-metal" thermal contact resistance, as well as the influence of metal diffusion into semiconductor and coefficient of charge collection by metal electrode on "TEM-metal" electrical contact resistance. On the other hand, from the experimental data [5-8] it is known that the quality of semiconductor surface treatment has a considerable impact on "TEM-metal" contact resistance, and, hence, on the quality of thermoelectric energy conversion. However, the author of this article is not familiar with the works which would logically, with regard to specific numerical characteristics of TEM surface roughness, consider its impact on the electrical resistance and thermoEMF of "TEM-metal" contact. Exactly this consideration is the purpose of the present study.

It should also be noted that in [9] the limits of the depth of the disturbed layer, which arises when cutting the ingot of thermoelectric material into legs, are given. It is believed that this depth can vary from 20 to 150 μm.

## 2. Calculation of the electrical contact resistance and thermoEMF of "TEM-metal" transient contact layer due to TEM surface roughness and discussion of the results obtained

We will first describe the physical model that was used in the calculation process. Let us have a TEM with a rough surface. Since such a surface can be imagined as a set of random "hollows" and "humps" with random depths and heights, we will draw imaginary horizontal planes through the "top" of the highest "hump" and through the "bottom" of the deepest "hollow". The distance between these parallel planes will be considered to be the known thickness $d_0$ of the transient layer. To "construct" the transient layer, we will fill with metal all the spaces between the horizontal planes free of TEM.

We now describe the methods of calculation of the electrical contact resistance and thermoEMF of "TEM-metal" transient contact layer in the framework of this physical model.

Let us start with the calculation of the contact resistance. Breaking contact area into elementary pads of size *ds*, we thereby break our transient contact layer into elementary bars of length $d_0$, interconnected in parallel. Each of them consists of a "metal" part of length $d_0 x$ and a "semiconductor" part of length $d_0(1-x)$, where $x$ – a random number from the range [0;1]. Therefore, the total conductivity of the contact is equal to:

$$\Sigma = \int_S \frac{ds}{d_0[\rho_m x + \rho_s(1-x)]}, \tag{1}$$

where $\rho_m$ and $\rho_s$ – resistivities of metal and TEM, respectively, $S$ – contact area. Applying mean value theorem to (1), we find the following resultant expression for the electrical contact resistance:

$$r_c = d_0 \left\langle \frac{1}{\rho_m x + \rho_s(1-x)} \right\rangle^{-1}, \tag{2}$$

where angular brackets mean averaging over rather large sequence of random (pseudorandom) numbers from the range [0;1]. Another model assumption we will make is that our pseudorandom numbers will be assumed to be distributed in the specified interval in accordance with the so-called "truncated Gaussian distribution", which we will present in the form:

$$f(x) = \frac{\exp\left[-(x-a)^2/2s^2\right]}{\int_0^1 \exp\left[-(x-a)^2/2s^2\right]dx}, \tag{3}$$

where $a$ and $s$ – certain parameters, and $0 \leq a \leq 1$. The boundary $\sigma \to \infty$ corresponds to uniform distribution for which $f(x) \equiv 1$. Then the contact resistance is definitely equal to:

$$R_c = d_0 \left[\int_0^1 \frac{f(x)dx}{\rho_m x + \rho_s(1-x)}\right]^{-1}. \tag{4}$$

Using formulae (3) and (4), we consider the impact of distribution parameters which characterize the quality of semiconductor surface treatment on the value of contact resistance. The result of calculations of contact resistance for bismuth telluride-nickel couple are given in Fig.1 for the thickness $d_0 = 20$ μm and the values of $a$ equal to 0.928, 0.5 and 0.072, respectively. In so doing, we assumed that $\rho_m$=8.7·10$^{-6}$Ohm·cm, and $\rho_s$=1.25·10$^{-3}$Ohm·cm.

From the figure it is seen that with increasing $s$, the electrical contact resistance rather quickly (already at $s = 20$) reaches the asymptotic value which corresponds to $s = \infty$, i.e. the uniform distribution of "hollows" and "hills" along the rough surface. This value is equal to:

$$R_c = \frac{d_0(\rho_s - \rho_m)}{\ln(\rho_s/\rho_m)}, \tag{5}$$

i.e. about 5·10$^{-7}$Ohm·cm$^2$.

Moreover, from the figure it is seen that at low values of $s$ the electrical contact resistance is the lower, the closer to unity is the value of $a$. The value of $s = 0$ at $a = 1$ corresponds to absolutely smooth surface, so it is clear that in this case $R_c = 0$ (in the figure this case is not shown).

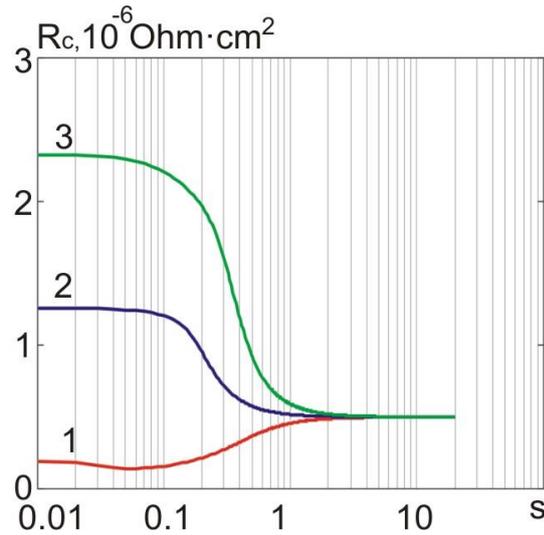

*Fig.1. Dependence of the electrical contact resistance of bismuth telluride-nickel at $d_0 = 20\mu m$ on the value of s at a equal to: 1) 0.928; 2)0.5; 3)0.072*

The figure also shows that, with decreasing *a*, the contact resistance increases, since a decrease in *a* corresponds to an increase in the length of the "semiconductor" and to a decrease in the length of the "metal" part of each elementary bar that forms the transient layer. Thus, according to curve 1, the lowest value of the contact resistance under the investigated conditions is about $2 \cdot 10^{-7}$ Ом·см$^2$.

The results of similar calculations for $d_0 = 150\,\mu m$ are given in Fig.2.

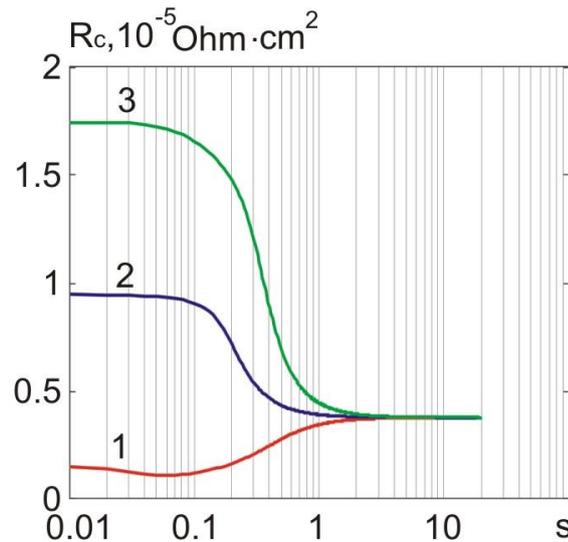

*Fig.2. Dependence of the electrical contact resistance of bismuth telluride-nickel couple at $d_0 = 150\mu m$ on the value of s at a equal to: 1) 0.928; 2)0.5; 3)0.072*

Under these conditions the lowest value of the contact resistance is $10^{-6}$Ohm·cm$^2$, and its asymptotic value is $3.75 \cdot 10^{-6}$Ohm·cm$^2$.

Thus, from the calculations it is seen that the character of surface treatment which is assigned by the distribution parameters (4) has a significant impact on the value of the contact resistance.

The best situation is realized when the difference in heights is significantly larger than surface roughness. Under these conditions, the "metal" part of the elementary bars is substantially larger than the "semiconductor", which explains the relatively low value of the contact resistance in this case.

Quite similarly, one can determine the thermal contact resistance due to surface roughness. It is given below:

$$R_t = d_0 \left[ \int_0^1 \frac{f(x)dx}{\kappa_m^{-1} x + \kappa_s^{-1}(1-x)} \right]^{-1}, \qquad (6)$$

The results of calculations of the thermal contact resistance by formula (6) are given in Figs.3, 4.

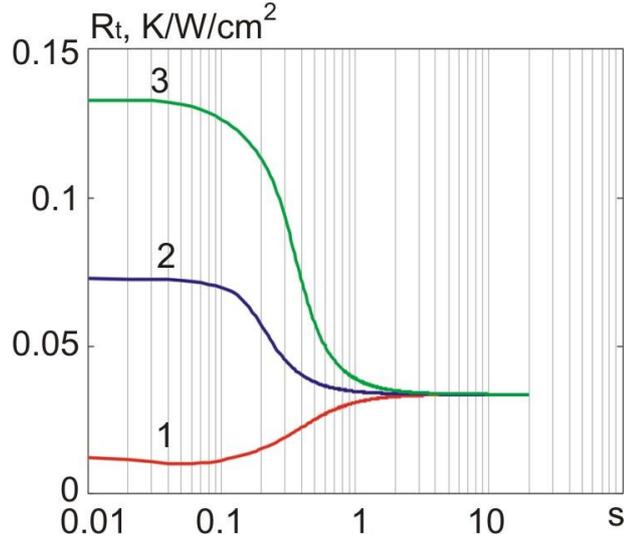

*Fig.3. Dependence of the thermal contact resistance of bismuth telluride-nickel couple at $d_0$ =20μm on the value of s at a equal to: 1) 0.928; 2)0.5; 3)0.072*

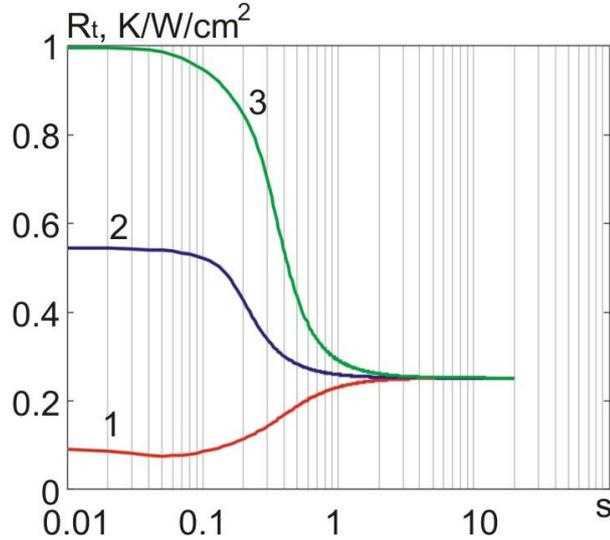

*Fig.4. Dependence of the thermal contact resistance of bismuth telluride-nickel couple at $d_0$ =150μm on the value of s at a equal to: 1) 0.928; 2)0.5; 3)0.072*

From the figures it is seen that the behavior of the thermal contact resistance as a function of *s* is quite similar to the behavior of the electrical contact resistance due to analogy between heat and charge transfer. The lowest value of the thermal contact resistance under conditions in question is equal to 0.01 K·cm²/W, and the asymptotic values at transient layer thicknesses 20 and 150μm are equal to 0.033 and 0.251 K·cm²/W, respectively.

The thermoEMF of the contact layer is found as the EMF of the parallel connected elementary bars, each of which has its own EMF and internal resistance, due to the ratio of lengths of the "metal" and "semiconductor" parts of each of them. Taking this into account, we first find the thermoEMF of an elementary bar. By definition, this thermoEMF is equal to the ratio of the difference of thermoelectric voltage on the bar to the temperature difference on it. So, first one must find the temperature distribution in the elementary bar. To this end, we write down the stationary equation of heat conductivity in the absence of external heat sources for the one-dimensional case. It will look like:

$$\frac{d}{dy}\left(\kappa \frac{dT}{dy}\right) = 0, \quad (7)$$

where $\kappa$ — coordinate-dependent thermal conductivity of the bar material. The general solution of this equation is as follows:

$$T = C_1 \int \frac{dy}{\kappa} + C_2, \qquad (8)$$

where $C_1, C_2$ – arbitrary constants that can be found from the initial conditions. Therefore, the thermoEMF of the elementary bar is:

$$\alpha_b = \frac{\int_0^{d_0} \alpha dT}{\int_0^{d_0} dT} = \frac{\int_0^{d_0}(\alpha/\kappa)dy}{\int_0^{d_0}(1/\kappa)dy} = \frac{(\alpha_m/\kappa_m)x + (\alpha_s/\kappa_s)(1-x)}{(1/\kappa_m)x + (1/\kappa_s)(1-x)}, \qquad (9)$$

where $\alpha_m, \alpha_s, \kappa_m, \kappa_s$ – the thermoEMF and thermal conductivities of metal and TEM, respectively.

Thus, the general thermoEMF of transient layer due to surface roughness is equal to:

$$\alpha_c = \frac{\int_0^1 [\rho_m x + \rho_s(1-x)]^{-1}[(\alpha_m/\kappa_m)x + (\alpha_s/\kappa_s)(1-x)][(1/\kappa_m)x + (1/\kappa_s)(1-x)]^{-1} f(x)dx}{\int_0^1 [\rho_m x + \rho_s(1-x)]^{-1} f(x)dx}. \qquad (10)$$

The results of calculation of the thermoEMF of bismuth telluride-nickel couple are given in Fig.5.

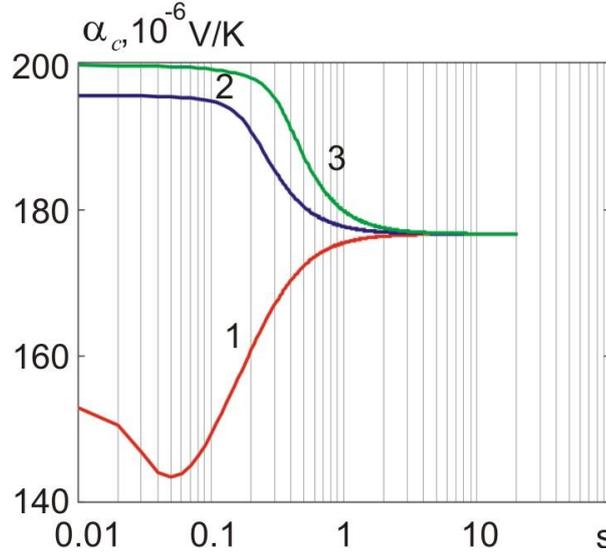

*Fig.5. Dependence of the thermoEMF of bismuth telluride-nickel couple at $d_0 = 150\mu m$ on the value of s at a equal to: 1) 0.928; 2)0.5; 3)0.072*

From the figure it is seen that the thermoEMF of bismuth telluride-nickel couple qualitatively depends on the value of *s* at different *a*, just as the electrical and thermal contact resistances. This similarity is due to the fact that as a result of the low thermal conductivity of TEM in comparison with the metal, other things being equal, the bulk of the temperature drop on the elementary bar falls on its semiconductor part. The lowest value of the thermoEMF in this case is about 145 μV / K, and its asymptotic value, which corresponds to the uniform distribution of "hollows" and "humps", is 176 μV / K. The case of "short circuit", when the transient layer is completely composed of metal, is not considered in this article.

**Conclusions**

1. It was established that the electrical contact resistance of "TEM-metal" transient layer due to semiconductor surface roughness for "nickel –bismuth telluride" couple at transient layer thickness 20μm is $5 \cdot 10^{-7} Ohm \cdot cm^2$, and its root-mean-square deviation is $1.13 \cdot 10^{-8} Ohm \cdot cm^2$.

2. It was established that the thermoEMF of "nickel-*p*-type bismuth telluride" transient contact layer due to semiconductor surface roughness, does not depend on contact layer thickness and is 23μV/K at practically zero root-mean-square deviation.

3. Since the value of the contact resistance that has to be "assigned" in order to match the real and estimated values of thermoelectric module parameters is an order of magnitude higher than that given in this article, this indicates, firstly, that the main part of the contact resistance is not due to semiconductor surface roughness, and secondly, that there are reserves for the contact resistance reduction. However, for their detection and use, further in-depth, including experimental, studies of the mechanisms of contact resistance formation are needed.